\documentclass[prc,english,twocolumn,superscriptaddress,floatfix]{revtex4}
\usepackage{babel}
\usepackage{ucs} 
\usepackage[utf8x]{inputenc}
\usepackage{amsmath}
\usepackage{amssymb}
\usepackage{amsfonts}
\usepackage[dvips]{graphicx}
\usepackage{verbatim}
\usepackage{hyperref}
\usepackage{subfigure}

\begin{document}
\title{{\em Ab initio} coupled-cluster theory for open-shell nuclei}
\date{\today}

\author{G.~R.~Jansen}

\affiliation{Department of Physics and Center of Mathematics for
  Applications, University of Oslo, N-0316 Oslo, Norway}

\author{M.~Hjorth-Jensen}

\affiliation{Department of Physics and Center of Mathematics for
  Applications, University of Oslo, N-0316 Oslo, Norway}

\author{G.~Hagen}

\affiliation{Physics Division, Oak Ridge National Laboratory, Oak
  Ridge, TN 37831, USA}
\affiliation{Department of Physics and Astronomy, University of
  Tennessee, Knoxville, TN 37996, USA}

\author{T.~Papenbrock}

\affiliation{Department of Physics and Astronomy, University of
  Tennessee, Knoxville, TN 37996, USA}

\affiliation{Physics Division, Oak Ridge National Laboratory, Oak
  Ridge, TN 37831, USA}

\affiliation {GSI Helmholtzzentrum f\"ur Schwerionenforschung GmbH,
  64291 Darmstadt, Germany}

\affiliation{Institut f\"ur Kernphysik, Technische Universit\"at
  Darmstadt, 64289 Darmstadt, Germany}

\begin{abstract}
  We develop a new method to describe properties of truly open-shell
  nuclei.  This method is based on single-reference coupled-cluster
  theory and the equation-of-motion method with extensions to nuclei
  with $A\pm 2$ nucleons outside a closed shell. We perform
  proof-of-principle calculations for the ground states of the helium
  isotopes $^{3-6}$He and the first excited $2^+$ state in $^6$He.
  The comparison with exact results from matrix diagonalization in
  small model spaces demonstrates the accuracy of the coupled-cluster
  methods.  Three-particle--one-hole excitations of $^4$He play an
  important role for the accurate description of $^6$He. For the
  open-shell nucleus $^6$He, the computational cost of the method is
  comparable with the coupled-cluster singles-and-doubles approximation
  while its accuracy is similar to coupled-cluster with singles,
  doubles and triples excitations.
\end{abstract}

\maketitle

\section{\label{sec:intro}Introduction}
The nuclear shell model is the paradigm for our understanding of
atomic nuclei~\cite{MayerJensen}. Within this model, doubly magic
nuclei (with fully occupied shells for protons and neutrons) are
particularly important because they are stronger bound than their
neighbors and can be approximated by a simple product of
single-particle states.  Within the nuclear shell model, doubly magic
nuclei are the cornerstones for our understanding of entire regions of
the nuclear chart as they can be viewed as inert cores. For the {\em
  ab initio} description of doubly magic nuclei, the coupled cluster  (CC)
method -- based on particle-hole excitations of a reference Slater
determinant that obey the linked cluster theorem -- is particularly
well suited and arguably one of the most efficient
methods~\cite{Coester1960,LKZ,shavittbartlett}.  Similar remarks hold
for nuclei that differ from doubly-magic nuclei by one nucleon; such
nuclei still exhibit a simple structure and a single Slater
determinant is a good reference state. The structure of all other
nuclei is more complicated and requires the superposition of many
product states and correspondingly large model spaces.

For the light $p$-shell nuclei, various {\em ab initio}
methods~\cite{Kamada2001, Varga1995, Barnea2000, Wiringa2001,
  Navratil2009, epelbaum2010a,epelbaum2010b} yield virtually exact
results for realistic Hamiltonians. For heavier systems, one typically
relies on approximations. Here, coupled-cluster theory is an ideal
compromise between accuracy on the one hand and computational cost on
the other.  This method has been applied to various problems in
nuclear
structure~\cite{Coester1960,LKZ,Guardiola,Heisenberg1999,Dean2004,Hagen2007c,Horoi2007,Hagen2008}.

In this paper, we use equation-of-motion (EOM) techniques within the
coupled-cluster method, abbreviated to EOM-CC hereafter, 
for the description of nuclei that differ from
closed-shell references by two nucleons. This
extension of the coupled-cluster method is useful for two reasons.
First, it significantly enlarges the set of nuclei that can be
accessed within coupled-cluster theory. For the oxygen isotopes
$^{14-28}$O, for instance, all nuclei except $^{19}$O differ from
closed-subshell references by two neutrons or less. Similar comments
apply to isotopes of helium and calcium. Questions related to the
evolution of shell structure~\cite{Sorlin,Kanungo2009} could thus be
addressed from first principles. Second, the coupled-cluster
method yields a similarity transformed Hamiltonian (see
Eq.~(\ref{eq:cc_barh}) below) for a doubly-magic nucleus. The
Hamiltonians for one and two nucleons attached to this doubly magic
core provide us with effective single-particle energies and an
effective two-body interaction, respectively. These matrix elements
could enter the construction of effective shell-model
interactions~\cite{HjorthJensen1995} that are the basis for large
scale shell-model calculations \cite{Brown2001, Honma2004}.

Within the EOM-CC methods, the equations for excited states and one
particle attached/removed are well known in quantum
chemistry~\cite{shavittbartlett,NooijenSnijders,NooijenBartlett,StantonGauss,Krylov,Piecuch},
and have also been applied to atomic nuclei
\cite{Wloch,Gour,Hagen2010,Hagen2010a}. However, the corresponding
equations for two particle attached/removed have seen very few
applications in quantum chemistry~\cite{NooijenBartl,Wladyslawski} and
are new in nuclear physics.  In this article, we will thus extend the
range of the EOM-CC methods to include open-shell nuclei with $A=\pm2$
nucleons outside a closed shell core. We present here the necessary
formalism for deriving such equations, including the pertinent
diagrams and algebraic equations.  To our knowledge, these details
have not been presented elsewhere.  The results from our EOM-CC
calculations are compared with full configuration interaction (FCI)
calculations for helium isotopes, demonstrating the accuracy
of this approach.

This paper is organized as follows. In Sec.~\ref{sec:EOMCC}, we give
a brief overview of the equation-of-motion method within coupled-cluster
theory.  The extension of this method to two valence nucleons outside
a closed-shell core is presented in Sec.~\ref{sec:2P-EOM-CC}. We
discuss and present our results in Sec.~\ref{sec:results}. 
Section \ref{sec:conclusion} contains our conclusions and an outlook for
future work.

\section{\label{sec:EOMCC}Coupled-cluster theory and
  equations-of-motion for nuclei}

\subsection{Single-reference coupled-cluster theory}
In this section we introduce the Hamiltonian that enters our
calculation, together with a brief review of single reference
coupled-cluster theory.  We keep the presentation limited to those
details that are required for the derivation of the two particle
attached/removed (2PA/PR-EOM-CC) amplitudes presented in
Sec.~\ref{sec:2P-EOM-CC}. The interested reader is referred to 
\cite{shavittbartlett} for details.

We use the intrinsic Hamiltonian
\begin{equation}
    \hat{H} = \left( 1 - \frac{1}{A} \right) \sum_{i=1}^A \frac{p_i^2}{2m} + \left[\sum_{i<j=1}^A \hat{v}_{ij} -\frac{\vec{p}_i \cdot \vec{p}_j}{mA}\right] \ . 
    \label{eq:hamilton}
\end{equation}
Here $A$ is the number of nucleons, and $\hat{v}_{ij}$ is 
the nucleon-nucleon interaction. We will limit ourselves to two-body 
interactions only. In second quantization, the Hamiltonian can be written as
\begin{equation}
\hat{H}=\sum_{pq} \varepsilon^p_q a^\dagger_p a_q 
+{1\over 4}\sum_{pqrs} \langle pq||rs\rangle a^\dagger_p a^\dagger_q a_s a_r \ .
\end{equation}
The term $\langle pq||rs\rangle$ is a shorthand for the matrix elements (integrals) of the two-body part of the Hamiltonian
of Eq.~(\ref{eq:hamilton}), $p,q,r$ and $s$ represent various single-particle states while $\varepsilon^p_q$ stands for the matrix elements of the one-body operator in Eq.~(\ref{eq:hamilton}). 
Finally, operators like $a^\dagger_q$ and $a_p$ create 
and annihilate a nucleon in the
state $q$ and $p$, respectively.  
These operators fulfill the canonical
anti-commutation relations.

In single-reference coupled-cluster theory, the many-body ground-state $|\Psi_0\rangle $ 
is given by the exponential ansatz,  
\begin{equation}
    |\Psi_0\rangle  = \exp{(\hat{T})} |\Phi_0\rangle \ .
    \label{eq:cc_exp_anz}
\end{equation}
Here, $|\Phi_0\rangle $ is a single-reference Slater determinant, 
and $\hat{T}$ is the cluster operator that generates correlations. The
operator $\hat{T}$ is expanded as a linear combination of
particle-hole excitation operators
\begin{equation}
    \hat{T} = \hat{T}_1 + \hat{T}_2 + \ldots + \hat{T}_A \, 
    \label{eq:cc_op_cluster}
\end{equation}
where $\hat{T}_n$ is the $n$-particle-$n$-hole excitation operator
\begin{equation}
  \hat{T}_n = \left(\frac{1}{n!}\right)^2 \sum_{a_\nu i_\nu}
  t_{i_1 \ldots i_n}^{a_1 \ldots a_n} a^\dagger_{a_1} \dots a^\dagger_{a_n} 
a_{i_n} \dots a_{i_1} \ .
    \label{eq:cc_op_npnh}
\end{equation}
We use throughout this work the convention that the indices $i j k\ldots$
denote states below the Fermi level (holes), while the indices
$a b c\ldots$ denote states above the Fermi level (particles). For an
unspecified state, the indices $pqr\ldots$ are used. The amplitudes
$t_{i_1 \ldots i_n}^{a_1 \ldots a_n}$ will be determined by solving
the coupled-cluster equations.  In the singles and doubles approximation we truncate 
the cluster operator as
\begin{equation}
    \hat{T} \approx \hat{T}_{\rm CCSD} \equiv \hat{T}_1 + \hat{T}_2 \ ,
    \label{eq:cc_op_ccsd}
\end{equation}
which defines the coupled-cluster approach with singles and doubles excitations, the so-called  
CCSD approximation. The unknown amplitudes result from the solution of the 
CCSD equations given by
\begin{eqnarray}
\label{eq:cc_eq_amp}
\langle \Phi_i^a|\bar{H}|\Phi_0\rangle &=& 0, \nonumber\\
\langle \Phi_{ij}^{ab}|\bar{H}|\Phi_0\rangle &=& 0 \ .
\end{eqnarray}
The term
\begin{equation}
    \bar{H} = \exp{(-\hat{T})}\hat{H}\exp{(\hat{T})} = \left( \hat{H} \exp{(\hat{T})} \right)_C,
    \label{eq:cc_barh}
\end{equation}
is the similarity transform of the normal-ordered Hamiltonian. The state
$|\Phi_{ij\dots}^{ab\dots}\rangle$ is a Slater determinant that differs
from the reference $|\Phi_0\rangle$ by holes
in the orbitals $ij\dots$ and by particles in the orbitals $ab\dots$. The
subscript $C$ indicates that only connected diagrams enter.

Once the $t^a_i$ and $t^{ab}_{ij}$ amplitudes have been determined
from Eq.~(\ref{eq:cc_eq_amp}), the correlated ground-state
energy is given by
\begin{equation}
  E_{\rm CC} = \langle\Phi_0|{\bar{H}}|\Phi_0\rangle + E_0 \ . 
    \label{eq:cc_eq_energy}
\end{equation}
Here, $E_0$ denotes the vacuum expectation value with respect to the reference
state. 
The coupled-cluster equations~(\ref{eq:cc_eq_amp}) show that the
reference state $|\Phi_0\rangle$ is an eigenstate of the similarity
transformed Hamiltonian~(\ref{eq:cc_barh}) within the space of 1$p$-1$h$
and 2$p$-2$h$ excitations.

\subsection{\label{sec:2P-EOM-CC}Equation-of-motion coupled-cluster theory for 
open-shell nuclei}
In this work, our focus is on the development of coupled-cluster
theory for truly open-shell nuclei, i.e., systems where no single
reference can be constructed without breaking symmetries (such
as rotational invariance).  One could apply the CCSD method to the
deformed (symmetry breaking) Hartree-Fock ground state of an open-shell
nucleus.  However, the restoration of angular momentum requires more
than singles and doubles cluster excitations (see for example 
Ref.~\cite{Hagen2007c}) 
and is computationally expensive. Here, we wish to stay within the
computationally inexpensive CCSD scheme.

Open-shell systems can be computed with multi-reference methods.  In
such an approach, many reference wave functions are included and
treated on an equal footing. However, the loss of mathematical simplicity
and transparency, and problems related to intruder states make these
multi-reference approaches difficult to pursue. For a detailed
discussion, we refer the reader to Ref.~\cite{Bartlett02}.
Equation-of-motion methods (see \cite{Krylov,Piecuch} for recent
reviews) avoid these problems as they exhibit the transparency and
computational simplicity of single-reference coupled-cluster theory.

Within the EOM-CCSD approach, the states of the $A\pm 2$ open-shell nuclei are
computed from the ground state of the $A$-body system as 
\begin{equation}
    |\Psi_\mu^{(A\pm 2)}\rangle = \hat{R}_\mu^{(A\pm 2)} |\Psi_0^{(A)}\rangle =  \hat{R}_\mu^{(A\pm 2)} \exp{(\hat{T})} |\Phi_0\rangle \ .
\end{equation}
Here, $\hat{R}_\mu^{(A\pm 2)}$ is a particle removal or particle
addition operator that generates an ($A \pm 2$)-body state from the
$A$-body coupled-cluster wave function. The label $\mu$ identifies the
quantum numbers (energy, angular momentum, ...) of the state of
interest.

The operator $\hat{R}_\mu$, and the energies $E_\mu$ of the states of interest solve the 
eigenvalue problem~\cite{shavittbartlett,NooijenSnijders,NooijenBartlett,StantonGauss,Krylov,Piecuch}
\begin{align}
    (\bar{H} \hat{R}_\mu^{(A \pm 2)})_C |\Phi_0\rangle  &= \omega_\mu \hat{R}_\mu^{(A \pm 2)} |\Phi_0\rangle \ .
    \label{EOM_master}
\end{align}
Here, the expression $(\bar{H} \hat{R}_\mu^{(A \pm 2)})_C$ denotes all terms that connect
the similarity transformed Hamiltonian $\bar{H}$ with the excitation 
operator $R_\mu^{(A\pm 2)}$. The energy difference $\omega_\mu\equiv E_\mu-E_0$ is the
excitation energy of the state $\mu$ in the nucleus $A\pm 2$ with respect to
the ground state of the reference nucleus with mass $A$.

The operators $\hat{R}_\mu$ relevant for this work are (we drop the
label $\mu$ for convenience)
\begin{eqnarray}
\label{eq:op_2pa}
\hat{R}^{(A+2)} = {1\over 2}\sum_{ba} r^{ab} a^\dagger_a a_b^\dagger 
+{1\over 6}\sum_{iabc}r_{i}^{abc} a^\dagger_a a^\dagger_b a^\dagger_c
a_i +\ldots \\
\label{eq:op_2pr}
\hat{R}^{(A-2)} = {1\over 2}\sum_{ij} r_{ij} a_i a_j
+{1\over 6}\sum_{ijka}r_{ijk}^{a} a^\dagger_a a_k a_j a_i +\ldots \ ,
\end{eqnarray}
where the unknown amplitudes $r$ (subscripts and superscripts dropped)
can be grouped into a vector that solves the eigenvalue problem of
Eq.~(\ref{EOM_master}).  The operator~(\ref{eq:op_2pa}) consists of a
$2p$-$0h$ term, a $3p$-$1h$ term, and in general up to an
$(A+2)p$-$Ah$ term. In this work, we will truncate the
operator~(\ref{eq:op_2pa}) at the $3p$-$1h$ level. Clearly, this
truncation will only be a good approximation for states in the $(A\pm
2)$-body system that have a relatively simple structure built on the
$A$-body nucleus. We will introduce two different truncations for the
particle attached and the particle removed method, and identify them
by the number of particle-hole excitations kept in the
operator~(\ref{eq:op_2pa}).  A truncation after the first term in
Eq.~(\ref{eq:op_2pa}) is referred to as 2PA-EOM-CCSD(2$p$-0$h$), while the
truncation after the second term is denoted as 2PA-EOM-CCSD(3$p$-1$h$).
Similarly for 2PR-EOM-CCSD, we will use the abbreviations
2PR-EOM-CCSD(0$p$-2$h$) and 2PR-EOM-CCSD(1$p$-3$h$) for truncations after the
first and second term in Eq.~(\ref{eq:op_2pr}), respectively. Table
\ref{tab:op_eom_2p} shows the excitation operators used in these
truncation schemes.

\begin{table}
\begin{ruledtabular}
\begin{tabular}{cc}
    Operator & Expression \\
    \hline
    $\hat{R}^{A+2}_{2p-0h}$ & $\frac{1}{2} \sum_{ab} r^{ab} a^\dagger_a a^\dagger_b$ \\
    $\hat{R}^{A+2}_{3p-1h}$ & $\frac{1}{2} \sum_{ab} r^{ab} a^\dagger_a a^\dagger_b +
        \frac{1}{6} \sum_{abci} r^{abc}_i a^\dagger_a a^\dagger_b a^\dagger_c a_i$ \\
    $\hat{R}^{A-2}_{0p-2h}$ & $\frac{1}{2} \sum_{ij} r_{ij} a_j a_i$ \\
    $\hat{R}^{A-2}_{1p-3h}$ & $\frac{1}{2} \sum_{ij} r_{ij} a_j a_i + \frac{1}{6} \sum_{aijk} r_{ijk}^a a^\dagger_a a_k a_j a_i$
\end{tabular}
\end{ruledtabular}
\caption{Definition of the EOM excitation operators for the two particles attached (removed)
  method, using a truncation at both the 2-particle-0-hole (0-particle-2-hole) and 
  3-particle-1-hole (1-particle-3-hole) level.
  The operator $\hat{R}^{A+2}_{2p-0h}$ defines 2PA-EOM-CCSD(2$p$-0$h$), the operator $\hat{R}^{A+2}_{3p-1h}$
  defines 2PA-EOM-CCSD(3$p$-1$h$), the operator $\hat{R}^{A-2}_{0p-2h}$ defines 2PR-EOM-CCSD(0$p$-2$h$)
  while the operator $\hat{R}^{A-2}_{1p-3h}$ defines 2PR-EOM-CCSD(1$p$-3$h$). 
These operators enter the eigenvalue problem of Eq.~(\ref{EOM_master}).}
    \label{tab:op_eom_2p}
\end{table}

We construct the matrix (i.e., the connected part of $\bar{H} \hat{R}$)
of the eigenvalue problem of Eq.~(\ref{EOM_master}) diagrammatically.  As
usual, lines directed upwards represent particle states, while lines
directed downwards represent hole states~\cite{shavittbartlett}. The
horizontal lines represent the operators and we use a heavy and a wiggly
line too differentiate the two operators in the composite diagrams.
Table \ref{tab:r_diag_elements} shows the diagrams corresponding to
the $r$ amplitudes, while the matrix elements of the 
similarity-transformed Hamiltonian are represented by the diagrams 
shown in Table~\ref{tab:barh_diag_elements}. These elements are well
known and computed from the corresponding contractions of the cluster
operator~(\ref{eq:cc_op_ccsd}) with the
Hamiltonian of Eq.~(\ref{eq:hamilton}) (see, for instance,
Ref.~\cite{shavittbartlett}). They result from the construction of the
operator~(\ref{eq:cc_barh}) after the CCSD
equations~(\ref{eq:cc_eq_amp}) have been solved.
\begin{table}
    \begin{ruledtabular}
\begin{tabular}{cc}
    Amplitude & Diagram \\
    \hline
    \\
    $r^{ab}$ &\includegraphics[scale=0.4]{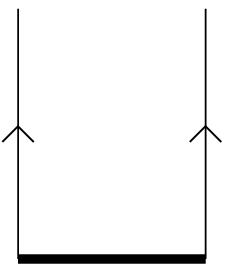} \\
    $r_{ij}$ & \includegraphics[scale=0.4]{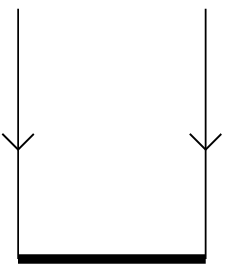} \\
    $r^{abc}_i$ & \includegraphics[scale=0.4]{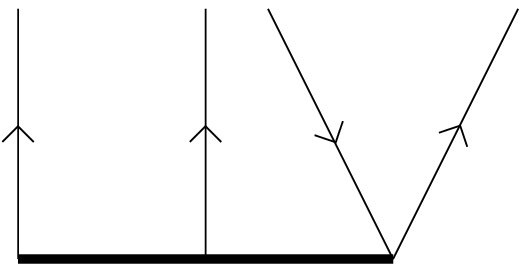} \\
    $r_{ijk}^a$ & \includegraphics[scale=0.4]{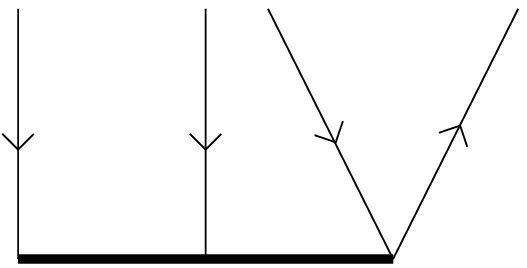} \\
\end{tabular}
\end{ruledtabular}
\caption{Diagrams corresponding to the excitation operators defined in Table~\ref{tab:op_eom_2p}.
  Upward directed lines denote unoccupied orbitals (particle states) and downward directed lines denote occupied orbitals (hole states). 
 The heavy horizontal line represents the 
operator vertex, used to distinguish the $\hat{R}$ operators from the 
        similarity transformed Hamiltonian $\bar{H}$.}
    \label{tab:r_diag_elements}
\end{table}
\begin{table}[ht]
    \begin{ruledtabular}
\begin{tabular}{cc}
    Matrix element & Diagram \\
    \hline
    \\
    $\langle i|\bar{H}|a\rangle$     & \includegraphics[scale=0.4]{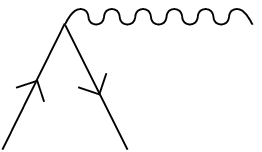} \\
    $\langle a |\bar{H}|b\rangle$     & \includegraphics[scale=0.4]{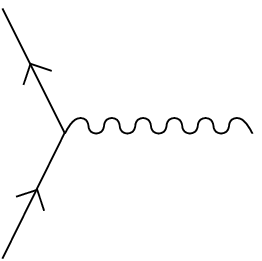} \\
    $\langle i |\bar{H}|j\rangle$     & \includegraphics[scale=0.4]{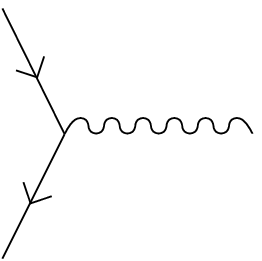} \\
    $\langle ai|\bar{H}|bc\rangle$   & \includegraphics[scale=0.4]{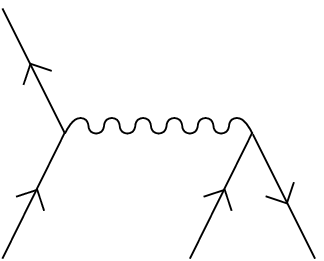} \\
    $\langle ij|\bar{H}|ka\rangle$   & \includegraphics[scale=0.4]{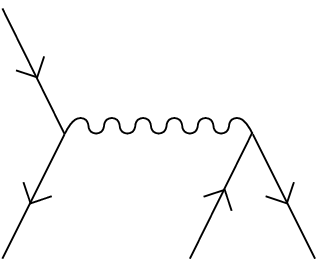} \\
    $\langle ab|\bar{H}|cd\rangle$   & \includegraphics[scale=0.4]{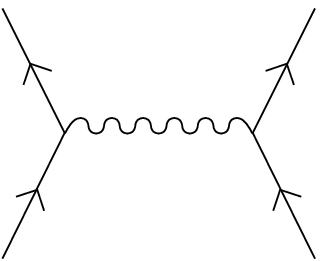} \\
    $\langle ij|\bar{H}|kl\rangle$   & \includegraphics[scale=0.4]{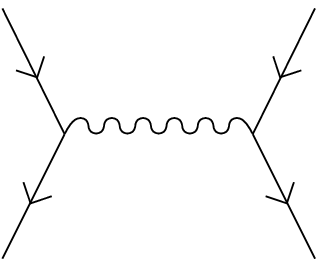} \\
    $\langle ia|\bar{H}|bj\rangle$   & \includegraphics[scale=0.4]{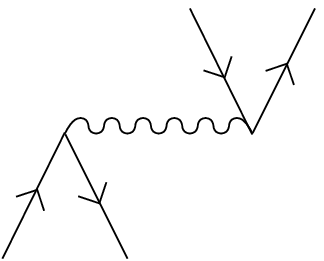} \\
    $\langle ab|\bar{H}|ci\rangle$   & \includegraphics[scale=0.4]{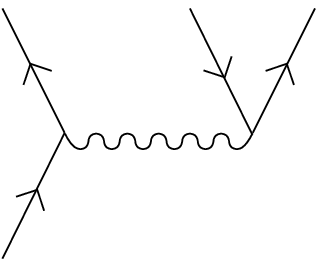} \\
    $\langle ia|\bar{H}|jk\rangle$   & \includegraphics[scale=0.4]{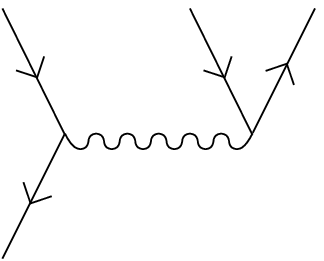} \\
    $\langle abc|\bar{H}|dei\rangle$ & \includegraphics[scale=0.4]{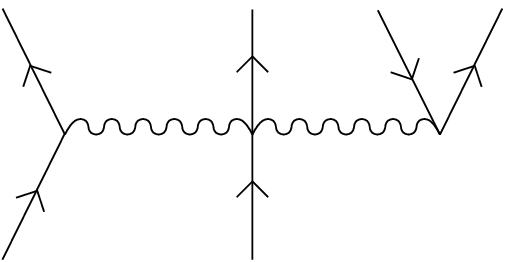} \\
    $\langle ija|\bar{H}|klm\rangle$ & \includegraphics[scale=0.4]{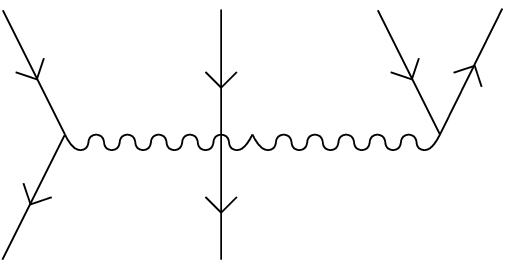} \\
    $\langle aib|\bar{H}|cdj\rangle$ & \includegraphics[scale=0.4]{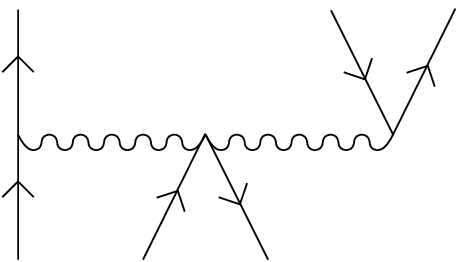} \\
    $\langle ija|\bar{H}|kbl\rangle$ & \includegraphics[scale=0.4]{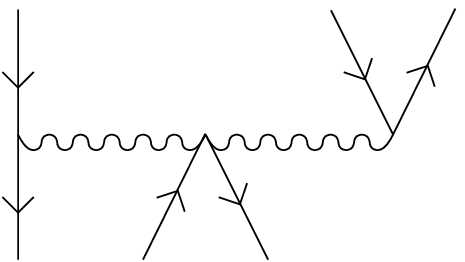} \\
\end{tabular}
\end{ruledtabular}
\caption{Diagrams of the matrix elements of the similarity-transformed Hamiltonian $\bar{H}$. The horizontal wiggly line represents the operator 
  vertex.
  The top three diagrams represent the one-body matrix elements of $\bar{H}$, the last 
  four diagrams represent the three-body matrix elements of $\bar{H}$, and the remaining 
  diagrams denote two-body matrix elements of $\bar{H}$, respectively. The corresponding 
  algebraic expressions 
  are shown in Table~\ref{tab:barh}. }
    \label{tab:barh_diag_elements}
\end{table}

Table \ref{tab:barh} shows the algebraic expressions for the matrix
elements of the similarity-transformed Hamiltonian $\bar{H}$. For
notational efficiency, some intermediate objects ($\chi$) that are 
common among several of the matrix elements, are defined separately in 
Table~\ref{tab:intermediates}.
In the numerical implementation, the storage of the similarity transformed
Hamiltonian requires some memory. However, this modest cost in memory yields a
significant reduction in computational cycles. For a detailed analysis
we refer the reader to Refs.~\cite{Kucharski,CCSD3NF}.
\begingroup
\squeezetable
\begin{table}
    \begin{ruledtabular}
\begin{tabular}{ccc}
    Matrix element & Shorthand & Expression \\
    \hline
    $\langle i|{\bar{H}}|a\rangle$ & $\bar{H}_{a}^i$ & $f_{a}^i + \langle im ||ae\rangle t_m^e$ \\
    $\langle a|{\bar{H}}|b\rangle$ & $\bar{H}_b^a$ &
        ${\chi}_b^a
        - \frac{1}{2} \langle {mn}||{be}\rangle t_{mn}^{ae}$ \\
        && $- t_m^a \bar{H}_{b}^m$ \\
    $\langle i |{\bar{H}}|j\rangle$ & $\bar{H}_j^i$ &
        $f_{j}^i$
        $+ \langle im ||je\rangle t_m^e$ \\
        &&
        $+ \frac{1}{2} \langle im ||ef\rangle t_{jm}^{ef}$
        $+ t_j^e \bar{H}_{e}^i$ \\
    $\langle ai|{\bar{H}}|bc\rangle$ & $\bar{H}_{bc}^{ai}$ &
        ${\chi}_{bc}^{ai}
        - \frac{1}{2} \langle {mi}||bc\rangle t_m^a$ \\
    $\langle ij|{\bar{H}}|ka\rangle$ & $\bar{H}^{ij}_{ka}$ &
        ${\chi}_{ka}^{ij}
        + \frac{1}{2} \langle ij ||ea\rangle t_k^e$ \\
    $\langle ab|{\bar{H}}|cd\rangle$ & $\bar{H}_{cd}^{ab}$ &
        $\langle {ab} ||cd\rangle
        + \frac{1}{2}\langle {mn}||cd\rangle t_{mn}^{ab}$ \\
        && $- \hat{P}(ab) t_m^b {\chi}^{am}_{cd}$ \\
    $\langle ij|{\bar{H}}|kl\rangle$ & $\bar{H}_{kl}^{ij}$ &
        $\langle ij ||kl\rangle
        + \frac{1}{2} \langle ij ||ef\rangle t_{kl}^{ef}$ \\
        && $+ \hat{P}(kl) t_l^e {\chi}_{ke}^{ij}$ \\
    $\langle ia|{\bar{H}}|bj\rangle$ & $\bar{H}_{bj}^{ia}$ &
        ${\chi}_{bj}^{ia}
        + \frac{1}{2}\langle {mi}||eb\rangle t_{jm}^{ae}$ \\
    $\langle ab|{\bar{H}}|ci\rangle $ & $\bar{H}_{ci}^{ab}$ &
        $\frac{1}{2}\langle {ab}||ce\rangle t_i^e
        + {\chi}_{ci}^{ab}
        - t_{mi}^{ab} \bar{H}_c^m$ \\
        && $- \frac{1}{2} t_{mn}^{ab} \bar{H}_{ic}^{mn}
        + \hat{P}(ab) t_{mi}^{eb} \bar{H}_{ce}^{am}$ \\
        && $- \hat{P}(ab) t_m^a {\chi'}_{ci}^{mb}$ \\
    $\langle ia|{\bar{H}}|jk\rangle$ & $\bar{H}_{jk}^{ia}$ &
        ${\chi}_{jk}^{ia}
        + t_{jk}^{ea} \bar{H}_e^i$ \\
        && $+\hat{P}(jk) t_{mk}^{ea} \bar{H}_{je}^{im}
        - \frac{1}{2} t_m^a \bar{H}_{jk}^{im}$ \\
    $\langle abc|{\bar{H}}|dei\rangle$ & $\bar{H}_{dei}^{abc}$ &
    $\hat{P}(a,bc) \langle {mn}||de\rangle t_m^a t_{ni}^{bc}$ \\
    && $- \hat{P}(a,bc) \langle am || de\rangle t_{mi}^{bc}$ \\
    $\langle ija|{\bar{H}}|klm\rangle$ & $\bar{H}_{klm}^{ija}$ &
    $\hat{P}(k,lm) \langle ij || ke\rangle t_{lm}^{ea}$ \\
    && $+ \hat{P}(k,lm) \langle ij || ef\rangle t_k^e t_{lm}^{fa}$ \\
    $\langle aib|{\bar{H}}|cdj\rangle$ & $\bar{H}_{cdj}^{aib}$ &
    $- \langle mi || cd \rangle t_{mj}^{ab}$ \\
    $\langle ija|{\bar{H}}|kbl\rangle$ & $\bar{H}_{kbl}^{ija}$ &
    $ \langle ij || eb \rangle t_{kl}^{ea}$
\end{tabular}
\end{ruledtabular}
    \caption{Algebraic expressions for the matrix elements of the similarity-transformed Hamiltonian in terms of the cluster 
amplitudes $t^a_i$ and $t^{ab}_{ij}$, the matrix elements $\langle ij || ef\rangle$  of the two-body interaction, and 
the one-body matrix elements $f_q^p = \varepsilon_q^p+ \sum_i \langle{pi} |v| qi \rangle$ of the normal-ordered Hamiltonian.
The permutation operator $\hat{P}_{pq}$ permutes the indices $p$ and $q$, and we define $\hat{P}(pq) = 1 - \hat{P}_{pq}$, 
$\hat{P}(pq,r) = 1 - \hat{P}_{pr} - \hat{P}_{qr}$, and $\hat{P}(p,qr) = 1 - \hat{P}_{pq} - \hat{P}_{pr}$.
The intermediates $\chi$ are defined in Table~\ref{tab:intermediates}.}
    \label{tab:barh}
\end{table}
\endgroup

\begingroup
\squeezetable
\begin{table}
    \begin{ruledtabular}
\begin{tabular}{cc}
    Intermediate  & Expression \\
    \hline
    ${\chi}_b^a$ & $f_b^a + \langle am ||be\rangle t_m^e$ \\
    ${\chi}_{bc}^{ai}$ &
        $\langle {ai}||bc\rangle
        - \frac{1}{2} \langle {mi}|| bc \rangle t_m^a$ \\
    ${\chi}_{ka}^{ij}$ &
        $\langle ij || ka\rangle
        + \frac{1}{2} \langle ij ||ea \rangle t_k^e$ \\
    ${\chi''}_{bj}^{ia}$ &
        $\langle {ia}||bj\rangle
        + \frac{1}{2}\langle {ai}||eb\rangle t_j^e$ \\
    ${\chi'}_{bj}^{ia}$ &
        ${\chi''}_{bj}^{ia}
        + \frac{1}{2} \langle {ai}||eb\rangle t_j^e
        - \frac{1}{2}t_m^a {\bar{H}}_{jb}^{mi}$ \\
    ${\chi}_{bj}^{ia}$ &
        ${\chi'}_{bj}^{ia}
        - \frac{1}{2} t_m^a {\bar{H}}_{jb}^{mi}
        + \frac{1}{2}\langle {mi}||eb\rangle t_{jm}^{ae}$ \\
    ${\chi}_{ci}^{ab}$ &
        $\langle {ab}||ci\rangle
        + \frac{1}{2} \langle {ab}||ce\rangle t_i^e$ \\
    ${\chi}_{jk}^{ia}$ &
        $\langle {ia}||jk\rangle
        + \frac{1}{2} \langle {ia}||ef\rangle t_{jk}^{ef}
        + P(jk) t_j^e {\chi''}_{ek}^{ia}
        - \frac{1}{2} t_m^a \bar{H}_{jk}^{im}$ \\
\end{tabular}
\end{ruledtabular}
\caption{The intermediates that enter the construction of the similarity-transformed Hamiltonian $\bar{H}$ in 
Table \ref{tab:barh}. All other terms are defined in 
Table \ref{tab:barh}. } 
    \label{tab:intermediates}
\end{table}
\endgroup

\begin{figure}
    \subfigure[]{
        \includegraphics[scale=0.5]{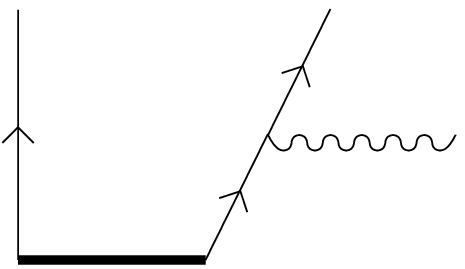}
        \label{fig:eom_diag_2pa_2p_1}
        }
    \subfigure[]{
        \includegraphics[scale=0.5]{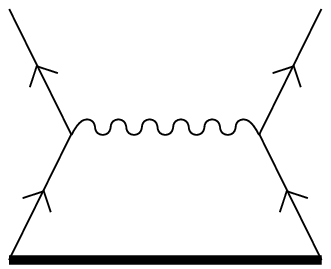}
        \label{fig:eom_diag_2pa_2p_2}
        }
        \caption{Diagrams corresponding to the matrix element $\langle
          \Phi^{ab} | (\bar{H}\hat{R})_C | \Phi_0 \rangle$ for the
          2PA-EOM-CCSD(2$p$-0$h$) amplitude equation. 
          All diagrams are
          constructed by contracting a diagram from Table \ref{tab:r_diag_elements}, with a diagram from 
          Table \ref{tab:barh_diag_elements}. Only diagrams that satisfy the topological form of 
          $\langle \Phi^{ab} | (\bar{H}\hat{R})_C | \Phi_0 \rangle$, with two external particle lines in 
          the upper part of the diagram and no external lines in the bottom part of the diagram, are selected.}
    \label{fig:eom_2pa_2p-0h}
\end{figure}

\begin{figure}
    \subfigure[]{
        \includegraphics[scale=0.5]{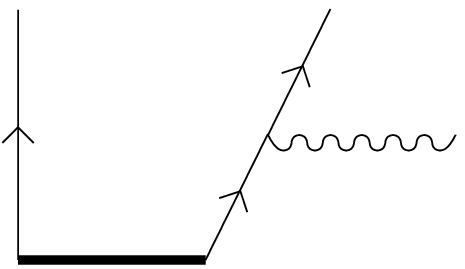}
        \label{fig:eom_diag_2pa_3p-1h_1}
        }
    \subfigure[]{
    \includegraphics[scale=0.5]{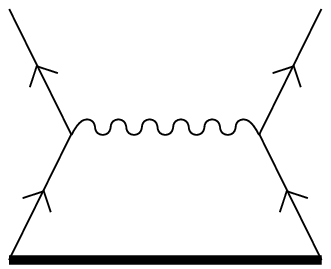}
        \label{fig:eom_diag_2pa_3p-1h_2}
        }
    \subfigure[]{
    \includegraphics[scale=0.5]{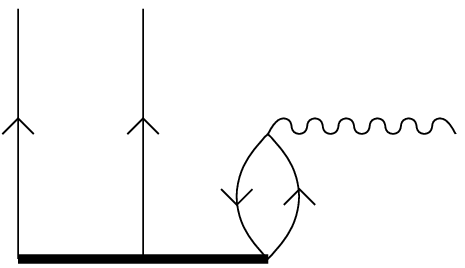}
        \label{fig:eom_diag_2pa_3p-1h_3}
        }
    \subfigure[]{
    \includegraphics[scale=0.5]{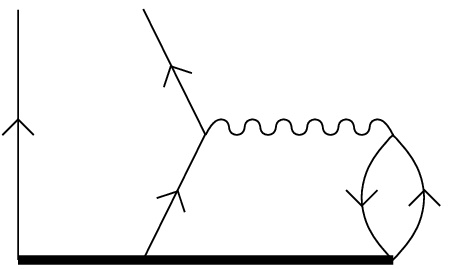}
        \label{fig:eom_diag_2pa_3p-1h_4}
        }
    \subfigure[]{
    \includegraphics[scale=0.5]{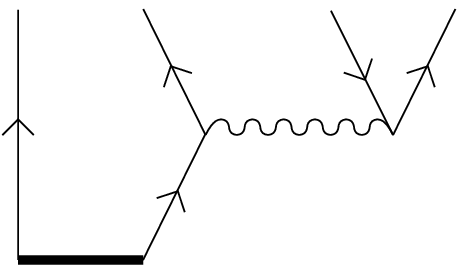}
        \label{fig:eom_diag_2pa_3p-1h_5}
        }
    \subfigure[]{
    \includegraphics[scale=0.5]{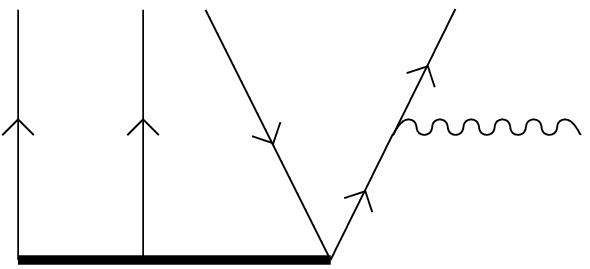}
        \label{fig:eom_diag_2pa_3p-1h_6}
        }
    \subfigure[]{
    \includegraphics[scale=0.5]{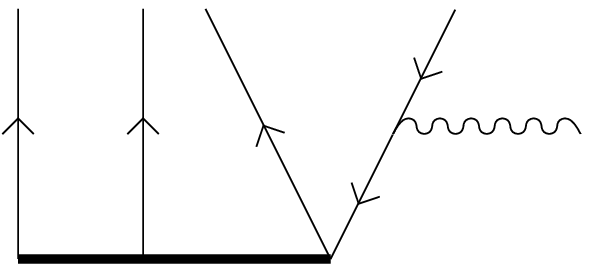}
        \label{fig:eom_diag_2pa_3p-1h_7}
        }
    \subfigure[]{
    \includegraphics[scale=0.5]{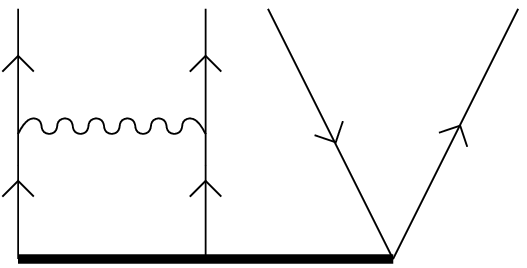}
        \label{fig:eom_diag_2pa_3p-1h_8}
        }
    \subfigure[]{
    \includegraphics[scale=0.5]{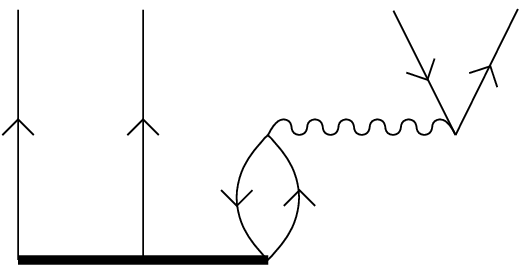}
        \label{fig:eom_diag_2pa_3p-1h_9}
        }
    \subfigure[]{
    \includegraphics[scale=0.5]{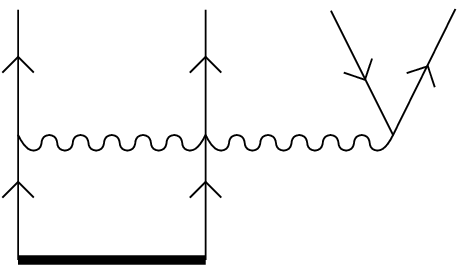}
        \label{fig:eom_diag_2pa_3p-1h_10}
        }
    \subfigure[]{
    \includegraphics[scale=0.5]{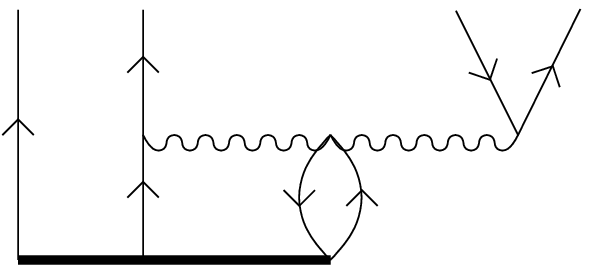}
        \label{fig:eom_diag_2pa_3p-1h_11}
        }
        \caption{Diagrams corresponding to the matrix elements 
          $\langle
          \Phi^{ab} | (\bar{H}\hat{R})_C | \Phi_0 \rangle$ (a-d) and 
          $\langle
          \Phi^{abc}_i | (\bar{H}\hat{R})_C | \Phi_0 \rangle$ (e-k)
          for the 2PA-EOM-CCSD(3$p$-1$h$) amplitude
          equation.}
    \label{fig:eom_2pa_3p-1h}
\end{figure}

In a diagrammatic language, the left-hand-side of the eigenvalue
problem~(\ref{EOM_master}) consists of all topologically different
diagrams that result from connecting a diagram from Table
\ref{tab:r_diag_elements} with a diagram in Table
\ref{tab:barh_diag_elements}.  Figures \ref{fig:eom_2pa_2p-0h} and
\ref{fig:eom_2pa_3p-1h} show the diagrams of
$(\bar{H} \hat{R})_C$ for the two truncations of the 2PA-EOM-CCSD
method. Let us briefly discuss some of these 
diagrams. 

For 2PA-EOM-CCSD(2$p$-0$h$), the relevant diagrams correspond to the
matrix element $\langle\Phi^{ab}|{(\bar{H} \hat{R})_C}|\Phi_0\rangle$,
i.e., they have two outgoing particle lines and consist of
contractions of the similarity transformed Hamiltonians with $r^{ab}$.
As an example, consider the diagram of
Fig.~\ref{fig:eom_diag_2pa_2p_2}. It results from contracting the
diagram of $r^{ab}$ (cf.  Table~\ref{tab:r_diag_elements}), with the
diagram of $\langle ab|\bar{H} | cd \rangle$ (cf.
Table~\ref{tab:barh_diag_elements}).  For 2PA-EOM-CCSD(3$p$-1$h$)
additional diagrams of the form $\langle\Phi^{ab}|{(\bar{H}
  \hat{R})_C}|\Phi_0\rangle$ enter because the amplitude $r^{abc}_i$
is also permitted.  In addition, diagrams corresponding to the matrix
element $\langle\Phi^{abc}_i|{(\bar{H} \hat{R})_C}|\Phi_0\rangle$
(i.e.,  diagrams with three outgoing particle lines and one incoming
hole line) enter. The diagram in Fig.~\ref{fig:eom_diag_2pa_3p-1h_8},
for instance, is constructed by contracting the diagram element
representing $r_i^{abc}$ in Table~\ref{tab:r_diag_elements}, with the
diagram element representing $\langle ab|\bar{H} | cd \rangle$ in
Table~\ref{tab:barh_diag_elements}.

\begin{figure}
    \subfigure[]{
    \includegraphics[scale=0.5]{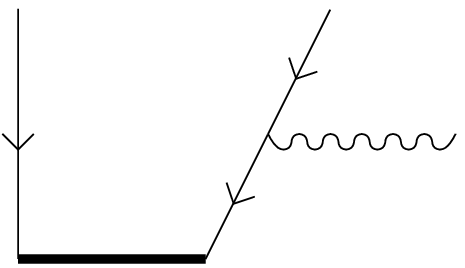}
        \label{fig:eom_diag_2pr_2h_1}
        }
    \subfigure[]{
    \includegraphics[scale=0.5]{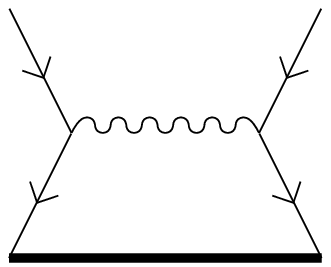}
        \label{fig:eom_diag_2pr_2h_2}
        }
        \caption{Diagrams corresponding to the matrix element $\langle
          \Phi_{ij} | (\bar{H}\hat{R})_C | \Phi_0 \rangle$ for the
          2PR-EOM-CCSD(0$p$-2$h$) amplitude equation.  All diagrams are
          constructed by contracting a diagram from Table \ref{tab:r_diag_elements}, with a diagram from 
          Table \ref{tab:barh_diag_elements}. Only diagrams that satisfy the topological form of 
          $\langle \Phi_{ij} | (\bar{H}\hat{R})_C | \Phi_0 \rangle$, with two external hole lines in 
          the upper part of the diagram and no external lines in the bottom part of the diagram, are selected.}
        \label{fig:eom_2pr_0p-2h}
\end{figure}

\begin{figure}
    \subfigure[]{
    \includegraphics[scale=0.5]{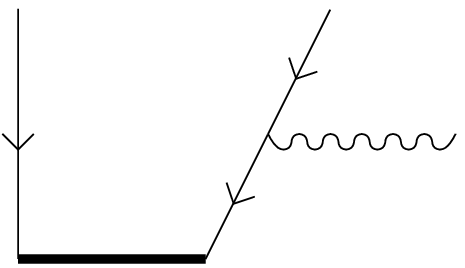}
        \label{fig:eom_diag_2pr_1p-3h_1}
        }
    \subfigure[]{
    \includegraphics[scale=0.5]{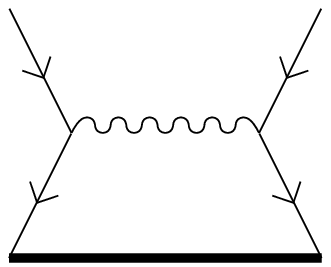}
        \label{fig:eom_diag_2pr_1p-3h_2}
      }
    \subfigure[]{
    \includegraphics[scale=0.5]{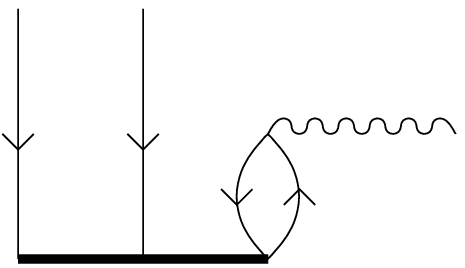}
        \label{fig:eom_diag_2pr_1p-3h_3}
        }
    \subfigure[]{
    \includegraphics[scale=0.5]{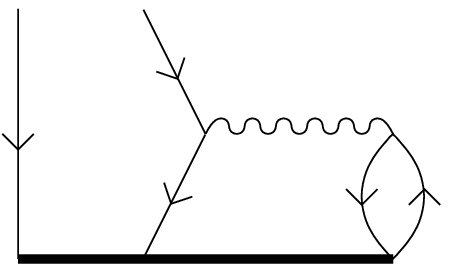}
        \label{fig:eom_diag_2pr_1p-3h_4}
        }
    \subfigure[]{
    \includegraphics[scale=0.5]{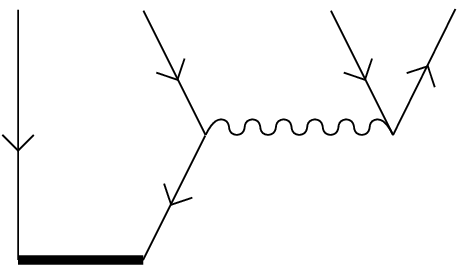}
        \label{fig:eom_diag_2pr_1p-3h_5}
        }
    \subfigure[]{
    \includegraphics[scale=0.5]{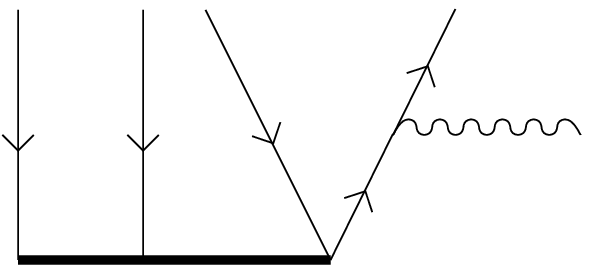}
        \label{fig:eom_diag_2pr_1p-3h_6}
        }
    \subfigure[]{
    \includegraphics[scale=0.5]{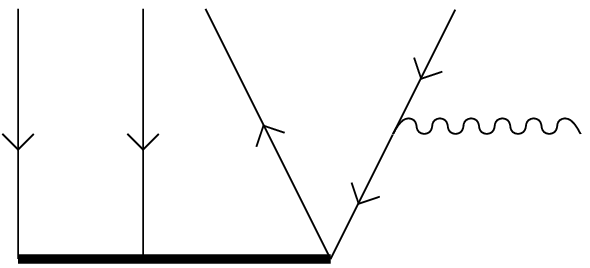}
        \label{fig:eom_diag_2pr_1p-3h_7}
        }
    \subfigure[]{
    \includegraphics[scale=0.5]{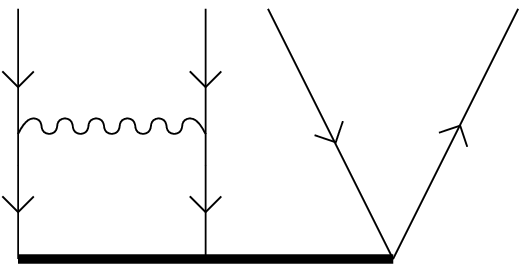}
        \label{fig:eom_diag_2pr_1p-3h_8}
        }
    \subfigure[]{
    \includegraphics[scale=0.5]{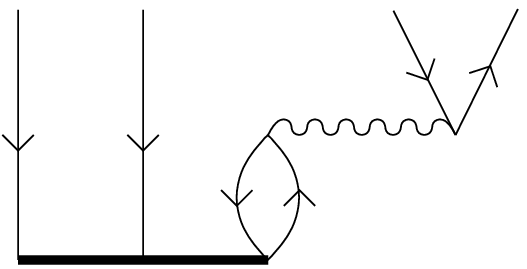}
        \label{fig:eom_diag_2pr_1p-3h_9}
        }
    \subfigure[]{
    \includegraphics[scale=0.5]{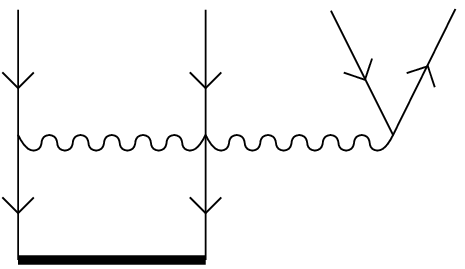}
        \label{fig:eom_diag_2pr_1p-3h_10}
        }
    \subfigure[]{
    \includegraphics[scale=0.5]{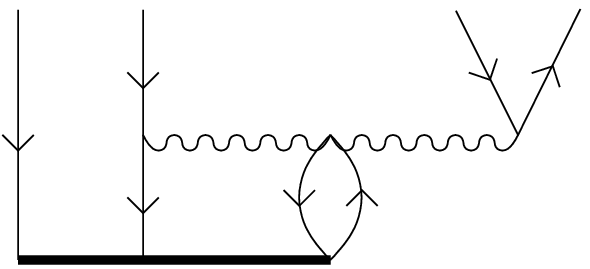}
        \label{fig:eom_diag_2pr_1p-3h_11}
        }
        \caption{Diagrams corresponding to the matrix elements 
          $\langle
          \Phi_{ij} | (\bar{H}\hat{R})_C | \Phi_0 \rangle$ (a-d) and 
          $\langle
          \Phi_{ijk}^a | (\bar{H}\hat{R})_C | \Phi_0 \rangle$ (e-k) for the
          2PR-EOM-CCSD(1$p$-3$h$) amplitude equation. 
  }
    \label{fig:eom_2pr_1p-3h}
\end{figure}

Let us turn to the diagrams for two-particle removal.
Figures \ref{fig:eom_2pr_0p-2h} and \ref{fig:eom_2pr_1p-3h} show
the diagrams of $(\bar{H} \hat{R})_C$ for the two truncations of the
2PR-EOM-CCSD method. Here, one needs topologies of the form
$\langle\Phi_{ij}|{(\bar{H} \hat{R})_C}|\Phi_0\rangle$ and
$\langle\Phi^{a}_{ijk}|{(\bar{H} \hat{R})_C}|\Phi_0\rangle$ using the
$r_{ij}$ and $r_{ijk}^a$ diagrams in Table \ref{tab:r_diag_elements}.

The algebraic expressions corresponding to these diagrams are derived
according to the standard rules, see for example
Ref.~\cite{shavittbartlett}, and shown in Table~\ref{tab:eq_eom}.  The
computational cost is $n_u^5n_o$ for the 2PA-EOM-CCSD(3$p$-1$h$)
method, and the most expensive diagram is shown in
Fig.~\ref{fig:eom_diag_2pa_3p-1h_8}.  Likewise, the most expensive diagram
for the 2PR-EOM-CCSD(1$p$-3$h$) is shown in
Fig.~\ref{fig:eom_diag_2pr_1p-3h_9} and requires of the order of
$n_u^2n_o^4$ operations. Here, $n_u$ is the number of
unoccupied orbitals (equal to the size of the valence space), and
$n_o$ is the number of occupied orbitals in the reference state. For
comparison, the computational costs of the single-reference CCSD and
CCSDT methods are $n_u^4n_o^2$ and $n_u^5n_o^3$, respectively. This
shows that the 2PA/2PR-EOM-CC methods developed in this work are
relatively inexpensive from a computational point of view, and
comparable to the cost of CCSD.  Note that the three-body matrix elements of the
similarity-transformed Hamiltonian are not stored. Instead, they are calculated when
needed and do not contribute significantly to the overall cost of
the calculations.  For a detailed analysis we refer the reader to
Refs.~\cite{Kucharski,CCSD3NF}.

\begin{table}
\begin{ruledtabular}
\begin{tabular}{cc}
    Matrix element & Expression \\
    \hline
    \multicolumn{2}{l}{\qquad2PA-EOM-CCSD(2$p$-0$h$):} \\
    \hline
    $\langle {\Phi^{ab}}| (\bar{H} \hat{R})_C |\Phi_0\rangle $ &
        $\hat{P}(ab) {\bar{H}}_e^b r^{ae}$
        $+ \frac{1}{2} \bar{H}_{ef}^{ab} r^{ef}$ \\
    \\
    \multicolumn{2}{l}{\qquad2PA-EOM-CCSD(3$p$-1$h$):} \\
    \hline
    $\langle {\Phi^{ab}}| (\bar{H} \hat{R})_C |\Phi_0\rangle $ &
        $\hat{P}(ab) {\bar{H}}_e^b r^{ae}$
        $+ \frac{1}{2} \bar{H}_{ef}^{ab} r^{ef}$ \\
        &
        $+ {\bar{H}}_{e}^m r_m^{abe}$
        $+ \frac{1}{2} \hat{P}(ab) {\bar{H}}_{ef}^{bm} r_m^{aef}$ \\
    $\langle {\Phi^{abc}_i}| (\bar{H} \hat{R})_C |\Phi_0\rangle $ &
        $\hat{P}(a,bc) {\bar{H}}_{ei}^{bc} r^{ae}$
        $+ \hat{P}(ab,c) {\bar{H}}_e^c r_i^{abe}$ \\
        &
        $- {\bar{H}}_i^m r_m^{abc}$
        $+ \frac{1}{2} \hat{P}(ab,c) {\bar{H}}_{ef}^{ab} r_i^{efc}$ \\
        &
        $+ \hat{P}(ab,c) {\bar{H}}_{ei}^{mc} r_m^{abe}$
        $+ \frac{1}{2} {\bar{H}}_{efi}^{abc} r^{ef}$ \\
        &
        $+ \frac{1}{2} \hat{P}(a,bc) {\bar{H}}^{bmc}_{efi} r_m^{aef}$ \\
    \\
    \multicolumn{2}{l}{\qquad2PR-EOM-CCSD(0$p$-2$h$):} \\
    \hline
    $\langle {\Phi_{ij}}| (\bar{H} \hat{R})_C |\Phi_0\rangle $ &
        $-\hat{P}(ij) {\bar{H}}_j^m r_{im}$
        $+ \frac{1}{2} {\bar{H}}_{ij}^{mn} r_{mn}$ \\
    \\
    \multicolumn{2}{l}{\qquad2PR-EOM-CCSD(1$p$-3$h$):} \\
    \hline
    $\langle {\Phi_{ij}} |(\bar{H} \hat{R})_C |\Phi_0\rangle $ &
        $-\hat{P}(ij) {\bar{H}}_j^m r_{im}$
        $+ \frac{1}{2} {\bar{H}}_{ij}^{mn} r_{mn}$ \\
        &
        $+ {\bar{H}1}_{e}^m r^e_{ijm}$
        $-\frac{1}{2} \hat{P}(ij) {\bar{H}}_{je}^{mn} r^e_{imn}$ \\
    $\langle {\Phi_{ijk}^a}| (\bar{H} \hat{R})_C |\Phi_0\rangle $ &
        $- \hat{P}(i,jk) {\bar{H}}_{jk}^{ma} r_{im}$
        $+ {\bar{H}}_e^a r^e_{ijk}$ \\
        &
        $- \hat{P}(ij,k) {\bar{H}}_k^m r^a_{ijm}$
        $+ \frac{1}{2} \hat{P}(ij,k) {\bar{H}}_{ij}^{mn} r^a_{mnk}$ \\
        &
        $+ \hat{P}(ij,k) {\bar{H}}_{ek}^{ma} r^e_{ijm}$
        $+ \frac{1}{2} {\bar{H}}_{ijk}^{mna} r_{mn}$ \\
        &
        $- \frac{1}{2} \hat{P}(i,jk) {\bar{H}}^{mna}_{jek} r_{imn}^e$
\end{tabular}
\end{ruledtabular}
\caption{Algebraic expressions for the 2PA/2PR-EOM-CCSD(2$p$-0$h$/0$p$-2$h$) and 2PA/2PR-EOM-CCSD(3$p$-1$h$/1$p$-3$h$) approximations. 
All terms are defined in Table~\ref{tab:barh}.}
\label{tab:eq_eom}
\end{table}

The eigenvalue problem of Eq.~(\ref{EOM_master}) for the two-particle attached system 
thus becomes
\begin{eqnarray}
\langle {\Phi^{ab}}| (\bar{H} \hat{R})_C |\Phi_0\rangle &=& \omega r^{ab}\nonumber\\
\langle {\Phi^{abc}_i}| (\bar{H} \hat{R})_C |\Phi_0\rangle &=& \omega r^{abc}_i \ .
\end{eqnarray}
Here, the left-hand-side is a linear function of the ``vector''
$R=(r^{ab},r^{abc}_i)$ of amplitudes (see Table~\ref{tab:eq_eom}) and
constitutes a matrix-vector product. Note that the two equations are
coupled and constitute a single eigenvalue problem for the 3$p$-1$h$
truncation. Likewise, we find for the two-particle removed problem
\begin{eqnarray}
\langle {\Phi_{ij}}| (\bar{H} \hat{R})_C |\Phi_0\rangle &=& \omega r_{ij}\nonumber\\
\langle {\Phi_{ijk}^a}| (\bar{H} \hat{R})_C |\Phi_0\rangle &=& \omega r^a_{ijk} \ .
\end{eqnarray}

We are usually only interested in the few lowest eigenvalues of
Eq.~(\ref{EOM_master}). For this purpose, we use the Arnoldi Method
for asymmetric eigenvalue problems, see for example
Ref.~\cite{golubvanloan} and references therein.  This method is based
on repeated applications of the matrix-vector product
$(\bar{H}\hat{R})_C$.  Specifically, our numerical implementation uses
the ARPACK software \cite{ARPACK} package.  The expressions in
Table~\ref{tab:eq_eom} can thus be used to solve the eigenvalue
problem directly.  In this paper, we employ the $m$-scheme basis for
the vectors $R=(r^{ab},r^{abc}_i)$ of the 2PA.  Within this scheme, we
are limited to small model spaces.  However, in this work we are
mainly interested in testing the newly developed methods and in
gauging the accuracy of the employed cluster truncation through
comparisons with exact diagonalization. Note that the
similarity-transformed Hamiltonian $\bar{H}$ exhibits the symmetries
of the underlying Hamiltonian. In the $m$-scheme basis, we can
classify states by their projections $J_z$ of angular momentum and
$T_z$ of isospin and their parity.  Although the solutions will have a
good total angular momentum $\hat{J}^2$, we will not be able to
exploit this symmetry in the $m$-scheme basis.

The FCI method we employ~\cite{factor} is limited to
relatively small model spaces.  Note, however, that we are also
working on an angular-momentum-coupled
implementation~\cite{gustav2011} of 2PA/2PR-EOM-CCM. This will allow
us to exploit rotational symmetries and give us access to much larger
model spaces.  These spaces are well beyond the reach of present full
diagonalization methods.

\section{\label{sec:results}Results}

For the proof-of-principle study we consider the helium isotopes
$^3$He to $^6$He. Here, $^3$He and $^5$He are viewed as one neutron
removed from and attached to $^4$He, respectively.  In a shell-model
picture, $^6$He is a truly open-shell nucleus with two valence
neutrons in the $p_{3/2}$ shell. Thus, a single-reference Slater
determinant may not be a good starting point for coupled-cluster
calculations of this nucleus, and it seems advantageous to describe
this six-nucleon system as two (halo) neutrons added to the $^4$He
core. We do not present results for $^2$He using the 2PR-EOM-CCSD
approach; an analysis of this method will be presented elsewhere
\cite{gustav2011}.

For our calculations we use a realistic nucleon-nucleon (NN) potential
derived from chiral effective field theory \cite{Weinberg1990288,
  Weinberg19913, Bedaque2002, Epelbaum2009, PhysRevC.68.041001} at
order N$^3$LO using interaction matrix elements from
Ref.~\cite{PhysRevC.68.041001}.  The matrix elements of the bare
interaction employ a cutoff at $\Lambda=500$ MeV. The short-range
parts of the interaction are removed via the similarity
renormalization group transformation \cite{PhysRevC.75.061001} with a
cutoff at $1.9$ fm${}^{-1}$. We use the spherical harmonic oscillator
with the oscillator frequency $\hbar\omega = 24$ MeV as our
single-particle basis. Our model space consists of five major
oscillator shells, with maximum orbital angular momentum $l_{\rm max}
= 2$.  This results in a total of 76 single-particle states for
neutrons and protons. We neglect three-body and four-body
interactions. This leads to missing many-body physics but is not
relevant for our proof-of-principle computation and comparison with
FCI calcualtions. 

We calculate the ground state energies of the A$=3$-$6$ helium isotopes.
For $^6$He we also compute the first $2^+$ state and the expectation
values of the total angular momentum.  In addition, we also discuss
the first $1^-$ excited state, as an example where the $3p$-$1h$
truncation fails.  We compare the equation-of-motion (EOM) approach to
the FCI method and to three different single-reference coupled-cluster
approximations. Recall that within the $m$-scheme, one can also
compute open-shell nuclei directly with the coupled-cluster method
without resorting to EOM techniques. While such a direct approach may
suffer from the lack of a good reference state (as is the case for the ground state
of $^6$He), the inclusion of more and more clusters must converge to
the FCI results.  Comparing the EOM-CC approach to these single-reference
coupled-cluster calculations allows us to gauge the efficiency of the
various coupled-cluster approximations.

The direct coupled-cluster calculations employ the CCSD approximation
described above, the CCSDT approximation (that includes triples
clusters) and the CCSDT-1 approach which includes some of the
$3p$-$3h$ clusters of the full CCSDT approximation.  For the EOM
calculations, we use the CCSD wave function of $^4$He as the
reference wave function and employ the intrinsic Hamiltonian of
Eq.~(\ref{eq:hamilton}) with $A=3,5,6$ for $^{3,5,6}$He, respectively.
We compare all results to an exact calculation done with the FCI
approach, using the same interaction and model space. The ground-state
energies of $^{3-5}$He are shown in Table~\ref{tab:he3he5}.  For
$^3$He, CCSDT becomes an exact method and agrees with FCI.  Here, the
single-reference coupled-cluster calculations are superior to the
EOM-CCSD approach. Evidently, the weakly bound nucleus $^3$He is not
well described as a neutron removed from the tightly bound $^4$He.
This means in turn that correlations beyond one-particle-two-hole
excitations play a non-negligible role.  For $^4$He, the EOM-CCSD approach is
identical to CCSD. Here, triples corrections and full triples
represent a significant improvement over CCSD, bringing the results
close to the FCI ones.  For $^5$He, the EOM-CCSD is superior to a
single-reference CCSD calculation and competes well with the
computationally more expensive triples correction CCSDT-1. Clearly,
the valence neutron in $^5$He is weakly correlated with the strongly
bound $^4$He core, and the PA-EOM-CCSD approach captures this
state very well.

\begingroup
\squeezetable
\begin{table}
\begin{ruledtabular}
\begin{tabular}{crrr}
                        & $^3$He   & $^4$He    & $^5$He \\ \hline
    CCSD                & $-6.624$ & $-27.468$ & $-22.997$  \\
    CCSDT-1             & $-6.829$ & $-27.600$ & $-23.381$  \\
    CCSDT               & $-6.911$ & $-27.619$ & $-23.474$ \\
    EOM-CCSD            & $-6.357$ & $-27.468$ & $-23.382$   \\
    FCI                 & $-6.911$ & $-27.640$ & $-23.640$  \\ \hline
\end{tabular}
\end{ruledtabular}
\caption{\label{tab:he3he5}Ground-state energies (in MeV) for $^3$He, $^4$He and $^5$He, 
  calculated with coupled-cluster methods truncated at the 2-particle-2-hole (CCSD) 
  level, 3-particle-3-hole (CCSDT) and a hybrid (CCSDT-1) 
  where a small subset of the leading diagrams in CCSDT are included.  
  For the EOM-CCSD approach, truncations has been made at the 1-particle-2-hole level, 
  the 2-particle-2-hole level, and the 2-particle-1-hole level
  for $^3$He, $^4$He and $^5$He respectively. The energies are compared 
  to the exact full configuration interaction (FCI).}
\end{table}
\endgroup

We turn to the truly open-shell nucleus $^6$He and show our results in
Table~\ref{tab:he6}. Here, the 2PA-EOM-CCSD (3$p$-1$h$) approach is
clearly superior to the single-reference coupled-cluster approaches,
as it reproduces the energy and the spin of the ground state to a very
good approximation. For the computation of the spin within the
single-reference approaches, we compute the expectation value $\langle
J^2\rangle$ within the Hellmann-Feynman theorem, and define $\langle
J\rangle$ from the relation $\langle J^2\rangle = \langle J\rangle
(\langle J\rangle +1)$. A direct computation within the CCSD or CCSDT
approximations can only be based on a symmetry-breaking reference
state.  Clearly, 2-particle-2-hole excitations (CCSD) cannot restore
the symmetry, and we find that $\langle J \rangle = 0.78$ for
the ground state of $^6$He. Adding 3-particle-3-hole excitations
(CCSDT approximation) almost restores the rotational symmetry, but
some correlation energy is still missing.  In the 2PA-EOM-CCSD
approach, however, the rotational symmetry is preserved throughout the
calculation, and we obtain a very good approximation of the energy at
a relatively low computational cost.  As expected, the 2PA-EOM-CCSD
(2$p$-0$h$) approach is less accurate than the $3p$-$1h$
approximation, since it is unable to account simultaneously for the
correlations within the three-body system consisting of the two
valence neutrons and the $^4$He core.  The ground state and
the first excited $2^+$ state of $^6$He are both dominated by a
configuration with two neutrons in the $p_{3/2}$ orbit. This is
consistent with the shell-model picture of this nucleus.
For the excited $2_1^+$ state, the CCSD, CCSDT and CCSDT-1 methods result in the 
correct value of the angular momentum due to the choice of reference state.

\begingroup
\squeezetable
\begin{table}
    \begin{ruledtabular}
\begin{tabular}{lllll}
    & $0_1^+$& $2_1^+$ &$0^+$ $\langle J \rangle$ &$2_1^+$ $\langle J\rangle$ \\
    \hline
    CCSD                     & $-22.732$  & $-20.905$ & $0.78$& 2 \\
    CCSDT-1                  & $-24.617$  & $-21.586$ & $0.25$& 2\\
    CCSDT                    & $-24.530$  & $-21.786$ & $0.01$& 2\\
    2PA-EOM-CCSD(2$p$-0$h$)  & $-21.185$  & $-18.996$ & $0$   & 2\\
    2PA-EOM-CCSD(3$p$-1$h$)  & $-24.543$  & $-21.634$ & $0$   & 2\\
    FCI                      & $-24.853$  & $-21.994$ & $0$   & 2\\
\end{tabular}
\end{ruledtabular}
\caption{\label{tab:he6}Energies (in MeV) for the ground state and first excited state of $^6$He 
  and the expectation value of the total angular momentum, calculated with coupled-cluster methods truncated at the 2-particle-2-hole (CCSD) level, 3-particle-3-hole (CCSDT) and a hybrid (CCSDT-1) where the 3-particle-3-hole amplitudes are treated perturbatively. 
  The 2PA-EOM-CCSD results are calculated with a truncation at the 2-particle-0-hole (2PA-EOM-CCSD(2$p$-0$h$) level and at the 3-particle-1-hole (2PA-EOM-CCSD(3$p$-1$h$) level.
  All energies are compared to full configuration interaction (FCI) results.}
\end{table}
\endgroup

Let us define the fraction $\sigma_{\rm corr}$ of the
correlation energy as
\begin{equation}
    \sigma_{\rm corr} = \frac{E_{\rm 2PA} - E_0}{E_{\rm FCI} - E_0} \ . \label{eq:accuracy}
\end{equation}
Here $E_{\rm 2PA}$ is the energy from 2PA-EOM-CCSD, $E_{\rm FCI}$ is
the energy from the FCI, while $E_0$ is the energy expectation of the
uncorrelated reference state $|\Phi_0 \rangle$.  Both $E_{2PA}$ and
$E_{\rm FCI}$ are shown in Table \ref{tab:he6}, and $E_0 =-16.807$~MeV.  
We also compute the norm
\begin{equation}
\label{norm}
{\cal N}=\sum_{ab} |r^{ab}|^2 +\sum_{abi} |r^{ab}_i|^2 \ . 
\end{equation}
The normalized squared weights
\begin{eqnarray}
\label{weights}
\rho_1^2&\equiv&{\cal N}^{-1}\sum_{ab} |r^{ab}|^2 \nonumber\\
\rho_2^2&\equiv&{\cal N}^{-1}\sum_{abi} |r^{ab}_i|^2 \ , 
\end{eqnarray}
fulfill $\rho_1^2+\rho_2^2=1$ and measure the importance of the
2$p$-0$h$ and 3$p$-1$h$ amplitudes, respectively.

Table \ref{tab:deltaecorr} shows the relative correlation energies
$\sigma_{\rm corr}$ and the relative weights $\rho_1$ and $\rho_2$ of
2PA-EOM-CCSD(3$p$-1$h$) for the three lowest states with quantum
numbers $0^+$, $1^-$ and $2^+$ of $^6$He, respectively. We see that
2PA-EOM-CCSD accounts for more than 90\% of the correlation energy for
the $0^+$ and $2^+$ states, and that most of the weight is carried by
the 2$p$-0$h$ amplitude. The $1^-$ state, however, is not very
accurately reproduced and much of the correlation energy is lacking.
For this state, the 2PA-EOM-CCSD(3$p$-1$h$) energy is
$E_{1^-}=-20.95$~MeV and deviates considerably from the full CI result
$E_{1^-}=-23.26$~MeV. Consistent with this picture is the large weight
carried by the 3$p$-1$h$ amplitudes. For a converged computation, one
would presumably also need to include $4p$-$2h$ or higher clusters.
Inspection shows that the $J^\pi=1^-$ state is dominated by two neutrons
in the $0p_{3/2}$ single-particle state, but with an additional
1-particle--1-hole excitation of either a proton or a neutron to the
$0p_{3/2}$ state or the $0p_{1/2}$ orbit. These 3$p$-1$h$
configurations are energetically favored compared with a configuration
with one neutron in the $0p_{3/2}$ state and one in the $0d_{5/2}$
state, a configuration which can also give a $1^-$ state.  This
explains why this state is dominated by the $r_i^{abc}$ amplitudes.
Note finally that we cannot expect a separation of the center-of-mass
motion from the intrinsic dynamics in the small model space we
considered~\cite{Hagen2010a,CoM}. Thus, the low-lying $1^-$ state
might also exhibit considerable admixtures with spurious
center-of-mass excitation.

\begingroup
\squeezetable
\begin{table}
    \begin{ruledtabular}
\begin{tabular}{cccc}
    & $\sigma_{\rm corr}$ & $\rho_1$ & $\rho_2$ \\
    \hline
    $0^+_1$ & $0.96$ & $0.84$ & $0.16$ \\
    $1_1^-$ & $0.64$ & $0.34$ & $0.66$ \\
    $2_1^+$ & $0.93$ & $0.81$ & $0.19$ \\
\end{tabular}
\end{ruledtabular}
\caption{\label{tab:deltaecorr} The relative correlation energy $\sigma_{\rm corr}$ defined in Eq.~(\ref{eq:accuracy}) for the lowest states with quantum numbers $J^\pi=0^+$, $1^-$ and $2^+$ in $^6$He, respectively. The relative weights $\rho_1$ and $\rho_2$ of the 2$p$-0$h$ and 3$p$-1$h$ amplitudes, respectively, are defined in Eqs.~(\ref{weights}). }
\end{table}
\endgroup

The results of this section show that states in open-shell
nuclei that exhibit a simple structure imposed on a correlated core
can be well described by the EOM method. Moreover, the EOM wave
function preserves the symmetries of the Hamiltonian and is
expected to be useful in the computation of matrix elements
besides the energy.  Taking into account the low computational cost of
the EOM-CC methods as compared to the full CCSDT approach, our EOM
approach is clearly well suited for these selected states.  For states
where more complicated particle-hole excitations are prominent, the
various truncation schemes discussed here are insufficient and we will
need additional correlations in the EOM operator to reach satisfactory
results.  The ground state of $^3$He and the first excited $1^-$ state
of $^6$He discussed above, are examples in case.

\section{\label{sec:conclusion}Conclusions and future perspectives.}

We have developed and implemented the equation-of-motion coupled-cluster
method and performed microscopic calculations of helium isotopes
with up to two valence nucleons outside the closed-shell alpha
particle.  The comparison with full configuration interaction
calculations shows that the equation-of-motion coupled-cluster methods
yield accurate results for sufficiently simple states. The open-shell
nucleus $^6$He, for instance, can be viewed and computed as two weakly
correlated neutrons attached to the correlated core of $^4$He.  The
proof-of-principle calculations were performed in a reduced model
space for a comparison with results from exact diagonalizations. We
are working on implementing our formalism for the two-particle
attached and removed equation-of-motion coupled-cluster methods in an
angular momentum coupled basis. This will allow us to employ much
larger model spaces.  With this improvement, the first-principles
computation of semi-magic nuclei, single-particle energies, and
effective two-particle interactions for the nuclear shell-model can be
addressed.

\begin{acknowledgments}
  We thank {\O}yvind Jensen for several discussions.  TP thanks the Institut
  f{\"u}r Kernphysik, Technische Universit{\"a}t Darmstadt, and the
  GSI Helmholtzzentrum f{\"u}r Schwerionenforschung for their
  hospitality. This work was supported in parts by the U.S.
  Department of Energy, under Grant Nos.\ DE-FG02-96ER40963
  (University of Tennessee), and DE-FC02-07ER41457 (UNEDF SciDAC), the
  Research Council of Norway, and the Alexander von Humboldt Stiftung.
  This research used computational resources of the Notur project in
  Norway and the National Center for Computational Sciences at Oak
  Ridge National Laboratory.
\end{acknowledgments}

\end{document}